# Lurking inferential monsters?
## Quantifying bias in non-experimental evaluations of school programs

This study examines whether unobserved factors substantially bias education evaluations that rely on the Conditional Independence Assumption. We add 14 new within-study comparisons to the literature, all from primary schools in England. Across these 14 studies, we generate 42 estimates of selection bias using a simple matching approach. A meta-analysis of the estimates suggests that the distribution of underlying bias is centered around zero. The mean absolute value of estimated bias is 0.03σ, and none of the 42 estimates are larger than 0.11σ. Results are similar for math, reading and writing outcomes. Overall, we find no evidence of substantial selection bias due to unobserved characteristics. These findings may not generalise easily to other settings or to more radical educational interventions, but they do suggest that non-experimental approaches could play a greater role than they currently do in generating reliable causal evidence for school education.

Ben Weidmann and Luke Miratrix
Harvard Graduate School of Education

**Acknowledgements**
We are particularly grateful to the UK Department for Education for access to the National Pupil Database. We also thank the Education Endowment Foundation for access to their RCT archive and to the Education Datalab for outstanding technical support. This work reflects the views of the authors, and none of the aforementioned organisations.

# 1. Introduction

School systems are awash with observational data. But using these data to provide answers to causal questions requires the brave assumption that once observable factors have been accounted for, there are no systematic differences between students who take part in an educational program (the treated) and those who don't (the controls). This 'Conditional Independence Assumption' (CIA) is brave in that it is often impossible to test.[1] As Edward Leamer noted, the trouble with non-experimental data is that "there is no formal way to know what inferential monsters lurk beyond our immediate field of vision" (Leamer, 1983, p. 39). The threat of such monsters – the possibility that students who are involved in programs differ in systematic, unobserved ways from other students – has resulted in widespread distrust of causal claims based on observational data in education (Gargani, 2010).

In response, researchers who want to generate empirical causal claims have increasingly focused their attention on Randomized Controlled Trials (RCTs), a method which ensures that treated and control groups are alike in all characteristics, on average. Crucially, this includes unmeasured factors such as the effort that parents put into their children's education. As a result, RCTs can generate highly credible causal estimates of program effectiveness.

But RCTs have costs and limitations. For example, trials often involve administering expensive and time-consuming tests that wouldn't otherwise be needed. RCTs also tend to rely on relatively small, non-random samples of students. This leads to imprecise estimates that may or may not generalise to broader populations (Bell, Olsen, Orr, & Stuart, 2016; Lortie-Forgues & Inglis, 2019). In short, while RCTs have clear methodological strengths, they are not perfect. This leaves open the possibility that observational studies could make a substantial contribution to our knowledge of 'what works' for different students. Such a contribution, however, hinges on the plausibility of the CIA in school settings.

---

[1] This assumption is variously known as: 'unconfounded treatment assignment'; 'no hidden bias'; 'no unobserved confounding'; 'ignorable treatment assignment'; and 'selection on observables' (e.g. Stuart & Green, 2008). In some other cases, the assumption is described as 'no omitted variable bias' and 'exogeneity' (e.g. King, Lucas, & Nielsen, 2017).

While there is no direct test for the CIA, in some instances it is possible to assess whether observational methods can replicate RCT results by performing a within-study comparison. The method, described in detail in Section 2, compares an experimental estimate (the 'causal benchmark') to an observational analysis that uses the same treatment group. What varies across the two estimates is how the control group is selected: the RCT uses control units that are chosen at random, while the observational study uses a set of non-randomly-selected comparison units (St. Clair, Cook, & Hallberg, 2014). The difference between estimates from the RCT and the observational study can be interpreted as a measure of selection bias (Wong, Valentine, & Miller-Bains, 2017).

This selection bias is the focus of our work. We seek to find out whether there are substantial differences between RCT and observational impact estimates of school programs. We concentrate on schools not only because of their inherent importance to society, but also because school systems routinely collect unusually rich census data. The existence of these data means that the potential benefit of non-experimental studies is large. Moreover, the small number of existing within-study comparisons in school settings suggest that the results from RCTs and non-experiments are typically close, suggesting that this may be a promising context for observational analyses.

Our study advances the literature in two primary ways. First, we present 14 new within-study comparisons. This substantially expands the empirical evidence on selection bias in school evaluations. We also diversify the literature by including a wide range of interventions – from teacher professional development programs to incorporating chess into the curriculum – and extending this area of scholarship beyond the United States.

Having multiple interventions in our dataset enables our second contribution: collectively analysing the *distribution* of selection bias in our context. We do this through meta-analysis. This approach is central to addressing our core research question: **what is the typical magnitude of selection bias in school evaluations that rely on the Conditional Independence Assumption?** Our meta-analysis complements existing scholarship, which typically focusses on individual within-study comparisons or qualitative reviews of studies.[2] Meta-analysing multiple within-study comparisons provides a more powerful test for the presence of selection bias and presents a clearer picture of its likely magnitude in school evaluations than individual studies can do alone. This is particularly valuable given the relatively limited power of individual within-study comparisons.

More broadly, we aim to draw attention to the value of analysing selection bias as a distribution, rather than focusing on individual estimates. While it is essential to keep generating new within-study comparisons, as the literature expands there is also a growing need for synthesis. Meta-analysis may be helpful in examining when and where selection bias due to unobserved variables is likely to be problematic.

Finally, a note about scope. To estimate selection bias we chose a simple, widely-used approach to non-experimental evaluation: 1-1 matching at the school level. This method, described in detail in section 4, was specified before we began generating results. However, our approach is only one of many possible ways to conduct an observational study. There remain numerous unanswered methodological questions about which approaches are most likely to minimise selection bias and replicate RCT results (Imbens, 2015; Smith & Todd, 2005). Future research should address these questions by exploring how the estimated distribution of bias changes as a function of various methodological choices. In this paper we have limited ourselves to a first order question about non-experimental evaluations in schools: is there evidence of substantial selection bias in analyses that use a simple analytical approach relying on the Conditional Independence Assumption?

---

[2] A notable recent exception is Chaplin et al. (2018) which examined the deviations between Regression Discontinuity Designs and Randomized Controlled Trials in 15 studies.

The paper proceeds as follows. Section 2 lays out the conceptual framework for our within-study comparisons, clarifying the strengths and limitations of this approach for ascertaining the validity of quasi-experimental methods. Sections 3 reviews existing scholarship on selection bias in school settings. Section 4 describes our unique dataset and the quasi-experimental method that we use in our analyses. Section 5 presents the main results, and a final section concludes.

## 2. Conceptual framework for within-study comparisons

Consider the following setup, based largely on Imai, King and Stuart (2008). Let the unit of analysis be schools and consider an observed sample of n units, drawn from a population of N schools. Let $S$ be a binary variable that indicates "sample selection" for schools. $S_j = 1$ when school $j$ takes part in a program evaluation (either through explicit agreement, or by not opting out). For an experimental study, $S_j$ is almost always non-random and is often defined by whether or not schools are interested in a particular intervention. They must also be willing to participate in the evaluation, which often includes administering tests. In observational studies of census data, $S_j$ equals 1 for almost all units. In other observational evaluations, $S_j$ can depend on factors such as testing regimes and data collection requirements in different locations.

Next, let $T_j$ be a binary variable indicating "treatment selection" i.e. whether a school is allocated to treatment or control. In the case of an RCT, $T_j$ is random by definition. In an observational study, $T_j$ reflects the decision school $j$ makes about whether it wants to implement a particular educational program. We follow the Neyman-Rubin causal model, where $Y_j(t)$ represents a potential outcome under treatment $t$. For us, this will be standardized test scores in reading, math and writing at the end of Grade 6 (when the average student is 11 years' old). Finally, let $X_j$ represent observed characteristics and $U_j$ represent unobserved factors.

The general set-up can be represented graphically as follows:

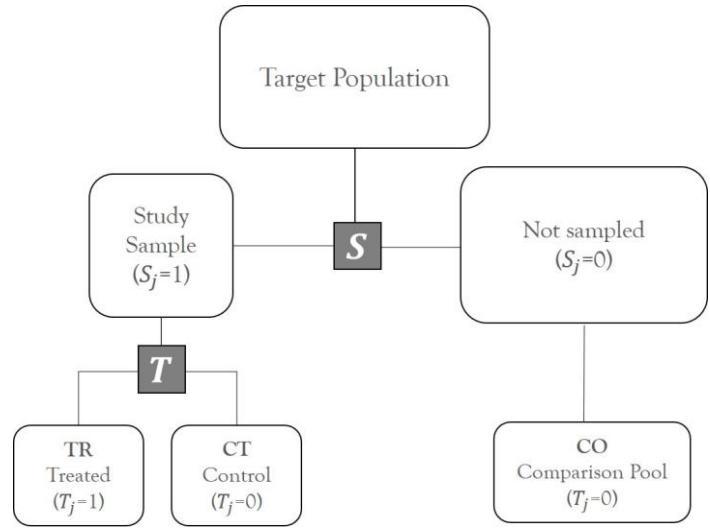

Within-study comparisons typically start by examining an RCT evaluation. The approach was devised by LaLonde (1986) who focused on a job-training experiment. In this canonical example, $S_j$ = 1 for people in the target population who met several conditions: first, living near one of the 10 job-training centres that agreed to use random assignment to determine program access; second, applying to the program; third, consenting to participate in research, including being randomised to receive job-training or no support. Within the group satisfying these criteria, participants were randomly assigned to treatment ($T_j = 1$) and control ($T_j = 0$).

Our analyses follow a similar pattern. $S_j = 1$ for school $j$ if three conditions are met: first, the school was located in an area where an intervention was being offered; second, the school decided that the intervention might be valuable to their students; and third, the school was willing to participate in an evaluation.

Let $\tau_{RCT}$ be defined as the average treatment effect for the treated (ATT) in the RCT:

$$\tau_{RCT} = E[Y(1)|T=1, S=1] - E[Y(0)|T=1, S=1]$$

The second expectation in this expression is not directly estimable, but randomisation means that an unbiased estimate of this quantity can be found using the outcomes of the RCT controls. As such $\hat{\tau}_{RCT}$ – our estimate of the treatment impact using an RCT – is a contrast between the observed outcomes in the treatment group (TR) and control group (CT). To illustrate, consider the simple difference-in-means estimator:

$$\hat{\tau}_{RCT} = \bar{Y}_{TR} - \bar{Y}_{CT}$$

LaLonde's insight was that this general setup also has the potential to yield a non-experimental estimate of the ATT ($\hat{\tau}_{Non-RCT}$). Instead of using the randomly selected control group (CT), analysts could use a non-randomly generated comparison group (CO) as a counterfactual. Focusing again on a difference-in-means for the sake of simplicity we have:

$$\hat{\tau}_{Non\_RCT} = \bar{Y}_{TR} - \bar{Y}_{CO}$$

$\hat{\tau}_{Non\_RCT}$ can be viewed as the ATT estimate that a researcher might have generated in the absence of randomisation. The difference between the non-experimental impact estimate ($\hat{\tau}_{Non-RCT}$) and the RCT estimate ($\hat{\tau}_{RCT}$) is then interpreted as an estimate of selection bias, denoted by $\hat{\beta}$:

$$\hat{\beta} = \hat{\tau}_{Non-RCT} - \hat{\tau}_{RCT}$$

As $\hat{\tau}_{RCT}$ and $\hat{\tau}_{Non-RCT}$ share the same treated units (TR), $\hat{\beta}$ can be simplified into a direct contrast between the observed outcomes in CO (the observational comparison group) and CT (the experimental control group; see Wong & Steiner, 2016). Specifically:

$$\hat{\beta} = \bar{Y}_{CT} - \bar{Y}_{CO}$$

It is worth emphasising that $\hat{\beta}$, our estimate of selection bias, focusses exclusively on outcomes under control. This is true for a vast majority of within-study comparisons, as the treatment group is typically shared across the RCT and observational estimates.[3]

$\beta$ can be thought of as comprising two main elements: bias associated with $U$ and bias associated with $X$ (Heckman, Ichimura, Smith, & Todd, 1998; Imai et al., 2008). To formalize this, let $\mu_{CO}(x) = E[Y(0)|X = x, T = 0, S = 0]$ be the mean comparison outcome, conditional on $X$. Defining $E\bar{Y}_{CO}^{adj}$ as the adjusted mean comparison outcome $E\bar{Y}_{CO}^{adj} = \int \mu_{CO}(x) dF_{CT}(x)$, we have:

$$\begin{aligned}
\beta &= E[\hat{\beta}] \\
&= E[\bar{Y}_{CT}] - E[\bar{Y}_{CO}] \\
&= E[\bar{Y}_{CT}] - E[\bar{Y}_{CO}^{adj}] + E[\bar{Y}_{CO}^{adj}] - E[\bar{Y}_{CO}] \\
&= \Delta_U + \Delta_X
\end{aligned}$$

$\Delta_X$ represents the contribution to bias made by differences in the values (or distribution) of $X$ across the control and comparison groups. $\Delta_U$ represents remaining bias after conditioning on $X$.

Provided there is overlap in the distribution of $X$ between the control and comparison groups, techniques such as matching or regression should minimise the contribution of $\Delta_X$ to $\beta$. Within-study comparisons therefore interpret non-zero estimates of $\beta$ as being evidence of $\Delta_U$ or 'hidden bias' (Cook, Shadish, & Wong, 2008). Large non-zero values of $\hat{\beta}$ are widely viewed as evidence that observational studies are unable to replicate RCT results, suggesting that the CIA is implausible in a particular setting.

---

[3] In rare cases, such as Shadish, Clark, and Steiner (2008), researchers have been able to use an independent within-study comparison, in which participants are randomly allocated to an RCT arm or an observational study arm (Wong, Steiner, & Anglin, 2018). In the RCT arm, participants are then randomized to treatment or control.

In the within-study comparison literature, estimates of $\hat{\beta}$ are taken to be an indicator of the internal validity of observational designs, i.e. selection bias as a result of $T$. We argue that within-study comparisons are also informative about selection bias as a result of $S_{RCT}$ (the sample selection mechanism of an RCT). Our reasoning is straightforward: $T$ is random in experimental evaluations and so any systematic differences between the CT and CO groups must be due to non-randomness in $S$.

That said, there is often a large degree of overlap between the sample selection mechanism for RCTs ($S_{RCT}$) and the treatment selection mechanism in typical observational studies ($T_{Non\_RCT}$). This is particularly true of the education programs we analyse here. In our setting, $S_{RCT}$ and $T_{Non\_RCT}$ both depend on whether school leaders decide that their students would benefit from an outside program to raise achievement. Moreover, 13 of the 14 interventions we study use a wait-list design. This means that all schools who choose to be part of an evaluation ($S_{RCT} = 1$) were guaranteed to receive the treatment at some point.[4]

As such, while $S$ represents 'selection into an RCT research sample' there are strong similarities with the process that governs $T$ in observational studies (i.e. $T_{Non\_RCT}$). It's therefore plausible that within-study comparisons in education RCTs can be informative about the trustworthiness of observational evaluations in schools.

## 3. Evidence from existing within-study comparisons

LaLonde's (1986) original paper was sceptical about the potential for non-experimental approaches to reproduce RCT estimates. Subsequent analyses, many of which followed LaLonde in focussing on job-training interventions, generally reinforced this scepticism. Three early reviews that examined the results of multiple within-study comparisons suggested that, while non-experimental estimates sometimes reproduced RCT treatment effects, they often produced results that differed substantially from RCT benchmarks (Bloom, Michalopoulos, & Hill, 2005; Glazerman, Levy, & Myers, 2003; Smith & Todd, 2005). These findings, which underscored the value of randomized designs, were influential in education policy circles (Wong et al., 2018).

---

[4] In two cases, the program offered to schools who were randomized to the control arm was a lighter-touch version of the program used in the main evaluation.

However, a later review of within-study comparisons by Cook, Shadish and Wong (2008) argued that there were circumstances in which observational studies were less prone to substantial selection bias. Cook, Shadish and Wong moved beyond the familiar territory of job-training programs and highlighted several factors. Non-experimental studies that succeeded in replicating trial results often had a high degree of overlap between treatment and comparison groups in terms of observed covariates. They also had access to covariates that were strong predictors of both selection and outcome.

More broadly, recent methodological work on within-study comparisons suggest that context is important in understanding the biasedness of non-experimental studies (Wong et al., 2018). The trustworthiness of non-randomized evaluations hinges on two context-dependent features: the quality of observational data available to analysts and the nature of the selection mechanism that determines treatment.

The setting we focus on is school education. A review of the literature revealed seven different interventions that had been examined using a within-study comparisons (summarised in Appendix A).[5] The cases analysed so far suggest that non-experimental estimators tend to perform well in education settings (Wong et al., 2017). Some of the best evidence comes from within-study comparisons of charter and 'magnet' school interventions (Abdulkadiroğlu, Angrist, Dynarski, Kane, & Pathak, 2011; Bifulco, 2012; Fortson, Verbitsky-Savitz, Kopa, & Gleason, 2012; Gill et al., 2016). Estimates of selection bias in this context are almost always smaller than $0.1\sigma$ (where $\sigma$ is shorthand for standard deviation and indicates that the scale for our analyses is effect-size units). These studies show no obvious pattern in the sign of selection bias estimates. This is true of both matching approaches and other analytical strategies.

---

[5] Other within-study comparisons in the field of education have focused on higher education (e.g. Angrist, Hudson, & Pallais, 2015; Pohl, Steiner, Eisermann, Soellner, & Cook, 2009; Shadish et al., 2008; Steiner, Cook, Shadish, & Clark, 2010) or pre-K (e.g. Dong & Lipsey, 2018). These contexts differ from school education in two important respects. First, they tend not to have access to the same detailed covariate information. Second, the target population of interventions are likely to differ. As Bifulco (2012) notes, school interventions – such as charter schools – which are typically focussed on disadvantaged and lower-achieving students, while universities are typically populated with higher achieving students.

While these results are promising in terms of the plausibility of the CIA, they don't necessarily apply to all evaluations in schools. In particular, they may not apply to school-based interventions and practices. Whether or not a child attends a charter school is a family decision, often made in the context of relatively disadvantaged neighbourhoods. In contrast, schools decide to adopt most educational programs, with these decisions being heavily influenced by principals and teachers. The unobserved factors shaping choices in these two different contexts may well differ.

Moving beyond charter and magnet schools, there is a lack of empirical work examining selection bias in school settings. Only three programs have been analysed thus far with a within-study comparison and two of these had limitations. Wilde and Hollister's (2007) research on the Tennessee STAR class size intervention did not have access to a pre-treatment outcome measure. Their mean estimate of $\hat{\beta}$ was $0.17\sigma$, with the researchers noting that site-level estimates of $\hat{\beta}$ were often very large (ranging from $-2.6\sigma$ to $1.9\sigma$). They concluded that propensity score matching did a poor job of replicating RCT estimates. Meanwhile, Fryer (2014) had issues with the balance and common support in his observational study. The intervention analysed by Fryer was intentionally implemented in the worst-performing primary schools in the study population, making it impossible to find non-experimental comparison schools that shared a common support.

The third school program analysed with a within-study comparison was Indiana's Diagnostic Assessment Intervention. Several papers focus on this project, which provides a pre-treatment outcome measure along with an observational comparison group that shared substantial overlap with the experimental sample (Hallberg, Wong, & Cook, 2016; St. Clair et al., 2014). The results suggest that matching approaches were generally successful in replicating RCT estimates. Interrupted-time series models also came close to replicating RCT estimates, although they were somewhat sensitive to model selection.

In summary, the literature currently contains too few examples to draw conclusions about plausibility of the CIA for school interventions. This scarcity of cases – combined with the lack of power of individual within-study comparisons – is the primary limitation of existing scholarship. Moreover, estimates of bias are often presented without confidence intervals, especially for the three school programs that have been analysed. The absence of uncertainty estimates, combined with the small number of cases, has made it difficult to provide quantitative syntheses across within-study comparisons. This is a significant barrier to assessing the CIA and quantifying the likely magnitude of selection bias in school settings.

## 4. Data and methods

This section describes the data and analytical methods we use. Appendix D presents an example case, the *Chess in Schools* intervention evaluated by Jerrim et al. (2016).

### 4.1 Data

Our analysis relies on a unique set of linked databases in England. The key data source is an archive of RCTs maintained by the Education Endowment Foundation (EEF). Many of the RCTs in the archive can be linked to the National Pupil Database (NPD), a census of publicly-funded schools and their pupils.[6] The NPD contains standardised achievement measures at multiple time points – kindergarten, grade 2, grade 6, and grade 11 – along with detailed information about students' backgrounds. The NPD also has information on where students live, which lets us incorporate local government datasets measuring neighbourhood characteristics such as crime rates (defined for areas of around 1,500 households each). Last, pupils can be linked to school surveys that capture information on funding, governance, staffing and so on. The covariates available for analysis are summarised in Table 1, with detailed descriptions in Appendix B.

---

[6] This represents over 90 percent of English school children (DfE, 2015).

These data provide an excellent basis for within-study comparisons. First and foremost, outcome data is comparable across the experimental and observational comparison groups: both sets of students sit the same standardised tests at the same time.[7]

**4.2 Interventions**

We perform within-study comparisons on 14 programs, all of which were initially evaluated with an RCT. We included all available RCTs from the EEF archive in which students had completed their 'Key Stage 2' national assessment after randomization.[8]

The overarching purpose of these interventions was to raise academic achievement. There was substantial diversity in how different programs pursued this mission. Some were directly targeted at achievement outcomes. For example, the "Affordable Maths" program provided grade 6 students with 1-to-1 tutoring online from math graduates in Sri Lanka and India. Other interventions were less direct. "Chess in Schools", for example, sought to raise math achievement by teaching chess to strengthening students' logical reasoning skills. The interventions were all implemented in English primary schools. See Table 2 for summary information.

The outcomes we examined were standardised academic achievement at the end of grade 6. We make use of the fact that at the end of primary school almost all students in England sit standardized tests in math, reading and writing. For each intervention and outcome, we generate two estimates of bias. The first is a naïve estimate, calculated as a simple difference in means between the experimental control group (CT) and the entire population of potential matches (CO). The magnitude of naïve bias $|\hat{\beta}^{Naive}|$ provides an initial indication of how big the issue of selection bias might be (Wong et al., 2018). Second, we estimate bias *after* conditioning on detailed set of covariates, including a pre-treatment measure of academic performance. The next two sections describe the process we use to generate this second estimate, $\hat{\beta}^{Match}$.

---

[7] This has been a serial problem in the literature, going back to LaLonde (1986).
[8] Initially, our data access only included primary school outcomes. We have subsequently been granted access to national assessments in high school, and will incorporate more interventions in future work. Note also that we also excluded two RCTs in the archive because they did not have school identifiers that could be matched to the National Pupil Database.

**4.3 Matching**

We specified our analytical approach before generating estimates of selection bias for our sample of interventions.[9] We use 1:1 matching at the school level. For each experimental control school we find a match in the comparison pool (i.e. government-funded schools in England that has not been part of an EEF trial). Our simple method was chosen for two reasons. First, while it doesn't make use of cutting-edge techniques, we wanted an approach that reflected the current state of practice for typical non-experimental evaluations of school programs. Second, it is computationally cheap; this is necessary for our inference procedure, described in Section 5.

Matching is done without replacement. We match on a Mahalanobis distance with a propensity-score calliper (Rosenbaum, 2010).[10] We take simple, automated steps to fit a propensity score model for each intervention, considering both interactions between covariates and flexible terms in a generalised additive model (see Appendix C for a description of this process). This procedure was designed to mimic the steps taken in typical observational studies.

All the pre-treatment variables listed in Table 1 are included in the Mahalanobis distance. Student-level covariates such as gender are turned into school-level variables by taking school-level means. Although the large number of variables in our Mahalanobis distance threatens to create a dimensionality issue, our access to a very large pool of potential matches – via the census information in the NPD – means that imbalance on observable characteristics is rarely an issue (see Table 3 in Appendix C).

---

[9] We used the *Chess in Schools* RCT as a test case where we initially examined several methods in an exploratory phase of work, before deciding to generate bias estimates with our simple 1:1 school-matching approach.
[10] In most trials, the median number of potential controls within the calliper for each treated school is over 1,000. In no trial is the median number of available controls within the calliper less than 190.

### 4.4 Estimating Bias

We estimate selection bias directly using a multilevel model. In our model student $i$ in school $j$ is exposed to intervention $w$, and tested on outcome $k \in$ (Math, Reading, Writing):

$$Y_{ijkw} = \alpha_j + \gamma X_{ij} + \beta_{kw}^{Match} S_j + \epsilon_{ijkw}$$
$$\alpha_j \sim N(\alpha_0, \sigma_\alpha^2)$$
$$\epsilon_{ijkw} \sim N(0, \sigma^2)$$

All the variables used in the matching (i.e. those described in Table 1) are included in $X$. We fit this model to all control schools in the RCT of intervention $w$ as well and all the students in the matched schools. The core parameter of interest for outcome $k$ and intervention $w$ is $\hat{\beta}_{kw}^{Match}$. As described in Section 2, $\hat{\beta}_{kw}^{Match}$ can be interpreted as an estimate of selection bias due to unobservable characteristics.

We look at all three outcomes for each intervention in an effort to increase our power to detect selection bias. Most of the interventions in the sample are explicitly focused on more than one domain of learning.[11] That said, for trials with a particular focus, such as Shared Maths, it may be the case that the unobservable characteristics associated with selection-into-treatment *only* affect mathematics outcomes (for example, they may be related to qualities of the math department that aren't present in other departments). To allay these concerns we performed a secondary, unplanned analysis in which we only examined primary outcomes. This produced extremely similar results to those presented in section 5. We now turn to those results.

---

[11] Changing Mindsets; Magic Breakfast; ReflectEd; Philosophy for Children; Mind The Gap; Hampshire Hundreds; Learner Response System; Dialogic Teaching

Table 1 – Summary of covariates

| Category | Label | Level | Description[a] | Source* |
|---|---|---|---|---|
| Student achievement | Achievement_grade 2 | Student | Average achievement in reading and math in Grade 2 | NPD (Key Stage Achievement) |
|  | Late | Student | = 1 if student sits standardized exam a year late | NPD (Key Stage Achievement) |
|  | Early | Student | = 1 if student sits standardized exam in a year earlier than expected | NPD (Key Stage Achievement) |
| Demographics | Age | Student | Age of student in months | NPD (Pupil Census) |
|  | Free school meals | Student | =1 if student currently gets free school meals | NPD (Pupil Census) |
|  | Gender | Student | = 1 if female | NPD (Pupil Census) |
| Rurality | Metro | Student | = 1 if student lives in metro area | NPD (Pupil Census) |
|  | Small_metro | Student | = 1 if student lives in small metro area | NPD (Pupil Census) |
|  | Rural | Student | = 1 if student lives in very rural area | NPD (Pupil Census) |
|  | Very rural | Student | = 1 if student lives in very rural area | NPD (Pupil Census) |
| School-level Achievement | School_academic_mean | School | Predicted achievement in reading and math in Grade 6 (pre-year) | Modelled (based on NPD) |
|  | School_academic_growth | School | Ave. annual change in academic achievement in Grade 6 (4 years prior to RCT) | Modelled (based on NPD) |
|  | School_grade_level_growth | School | Ave. annual change in percent of Grade 6 at grade level (4 years prior to RCT) | Modelled (based on NPD) |
| School size and type | Voluntary_school | School | = 1 if school is a voluntary school (state-funded, often religious) | NPD (School census) |
|  | Academy_sponsor | School | = 1 if school is a sponsored academy | NPD (School census) |
|  | Academy_converter | School | = 1 if school is a converted academy | NPD (School census) |
|  | Other_type | School | = 1 if school type is not described by the types listed above | NPD (School census) |
|  | Ofsted | School | Integer values of 1 (outstanding) to 4 (inadequate) | Ofsted |
|  | School size | School | Total number of students in school in pre-year | NPD (Finance) |
|  | Type_secondary | School | = 1 if secondary school | NPD (School census) |
|  | Type_middle | School | = 1 if school is a middle school | NPD (School census) |
|  | Type_both | School | = 1 if school has primary and high school | NPD (School census) |
| Budget | Income | School | Total income in pre-year | NPD (Finance) |
|  | Outside budget | School | Pounds spent on outside programs, services, and ICT | NPD (Finance) |
| Staffing | TA Percent | School | Proportion of staff who are Teacher Assistants | NPD (Workforce) |
|  | Teacher pupil ratio | School | Pupil teacher ratio in pre-year | NPD (Workforce) |
| Location variables | Crime | LSOA* | Index of crime | English Indices of Deprivation |
|  | Housing | LSOA* | Index of housing quality | English Indices of Deprivation |
|  | IDACI* | LSOA* | Omnibus index of disadvantage | English Indices of Deprivation |

*NPD = National Pupil Database; IDACI = Income Deprivation Affecting Children Index; LSOA = Lower Super Output Area (census region). See Appendix B for details. Pre-year is the year before the RCT randomisation.

Table 2 – Summary of interventions

| Intervention | ID | Brief description of intervention | n_schools* (n_pupils) | Reference |
|---|---|---|---|---|
| Affordable Online Maths Tuition | am | 1-on-1 online tutoring, for grade 6's by math graduates in India and Sri Lanka. ~45 mins per week for 27 weeks. | 64 (3,106) | Torgerson et al. (2016) |
| Changing Mindsets | cm | Professional development course for primary school teachers in how to develop Growth Mindset in pupils. | 30 (1,505) | Rienzo et al. (2015) |
| Chess in Schools | chs | Grade 5 students taught chess by experienced chess tutor, instead of music or PE, over 30 weeks. | 100 (4,009) | Jerrim et al. (2016) |
| Dialogic Teaching | dt | Grade 5 teachers trained in techniques to encourage dialogue, argument and oral explanation during class time | 78 (4,958) | Jay et al. (2017) |
| Flipped Learning | Fl | Grade 5 pupils learn core math content online, outside of class time. Classes were used to reinforce/clarify ideas. | 24 (1,214) | Rudd, Aguilera, Elliot, and Chambers (2017) |
| Hampshire Hundreds | hh | Professional development for primary schools teachers in strategies to reduce educational achievement gaps. | 36 (2,048) | McNally et al. (2014) |
| Learner Response System | lrs | Handheld devices used in grades 5 and 6, to provide teachers with real-time information about pupil knowledge | 97 (3,213) | Wiggins, Sawtell, and Jerrim (2017) |
| Magic Breakfast | mb | Providing nutritious breakfast to primary school students for most of the 2014-15 academic year. | 98 (4,038) | Crawford et al. (2016) |
| Mind the Gap | mtg | Teacher training and parent workshops (over a 5 week period) to help grade 4 students be more 'meta-cognitive'. | 45** (1,603) | Dorsett et al. (2014) |
| Philosophy for Children | p4c | Dialogic teaching of philosophical issues to children in grades 4 and 5, over a period of 11 months. | 48 (1,529) | Gorard et al. (2015) |
| ReflectEd | ref | Weekly lessons where grade 5's learn strategies to monitor/manage their own learning (over 6 months) | 33 (1,858) | Motteram et al. (2016) |
| Shared Maths | sm | Cross-age peer math tutoring: older pupils (grade 6) work with younger ones (grade 4) for 20 mins per week for 2 years. | 82 (3,167) | Lloyd et al. (2015) |
| Talk of the Town | tott | Whole-school intervention to help support the development of children's speech, language and communication. | 64 (3,299) | Thurston et al. (2016) |
| Thinking, Doing, Talking Science | ttds | 5 day's professional development for grade 5 teachers, with the aim of making science more practical and engaging. | 42 (1,513) | Hanley et al. (2015) |

*n_schools (pupils) describes the number of schools and pupils included in the original RCT evaluations at randomization.
**Figures based on the EEF Archive, rather than the published report, as the latter did not include the number of students at randomization.

## 5. Results

### 5.1 Bias Estimates

Raw estimates of selection bias are presented in Figure 1. The figure includes both $\hat{\beta}^{Match}$ (with 95 percent confidence intervals) and point estimates of $\hat{\beta}^{Naive}$. The boxplots at the bottom of the panel summarize the spread of $\hat{\beta}^{Naive}$ ($\Delta_X + \Delta_U$) and $\hat{\beta}^{Match}$ ($\Delta_U$).

These boxplots clearly illustrate the value of conditioning on covariates. The mean value of $\hat{\beta}^{Naive}$ is $-0.15\sigma$ suggesting that, in the absence of statistical controls, observational evaluations would typically suffer from negative bias and wrongly suggest that programs were harmful. In contrast, raw estimates of $\hat{\beta}^{Match}$ are centered very close to zero with a mean value of $-0.01\sigma$.

Two other features of the $\hat{\beta}^{Match}$ estimates stand out. First, 37 out of the 42 estimates of $\hat{\beta}^{Match}$ have a magnitude of less than $0.1\sigma$. This is broadly in line with previous within-study comparisons conducted in school settings. Second, very few estimates of $\hat{\beta}^{Match}$ (3 out of 42) are significantly different from zero ($\alpha$=0.05). This is arguably unsurprising. The RCTs that form the basis of our sample were generally powered to detect effect sizes of around $0.2\sigma$ and sometimes larger. Given our simple 1:1 matching approach, the power of the underlying RCT to detect an effect is a limit on the power of the within-study comparison to detect bias. Consequently, substantial selection bias due to unobserved characteristics could go undetected by individual within-study comparisons. This is the motivation for focussing on the distribution of bias, rather than noisy estimates from individual studies.

### 5.2 What Is the Typical Magnitude Of Selection Bias Due To Unobserved Differences?

At first glance, readers may be tempted to view the boxplots in Figure 1 as the distributions of selection bias before conditioning on observables ($\widehat{\boldsymbol{\beta}}^{Naive}$) and afterwards ($\widehat{\boldsymbol{\beta}}^{Match}$). However, these distributions reflect both bias and sampling variation. If unobserved covariates were pure noise, estimates of $\widehat{\boldsymbol{\beta}}^{Match}$ would still be non-zero due to sampling error. As such, the boxplots at the bottom of Figure 1 present a potentially misleading picture about the typical magnitude of bias, and how much it varies across studies and outcomes.

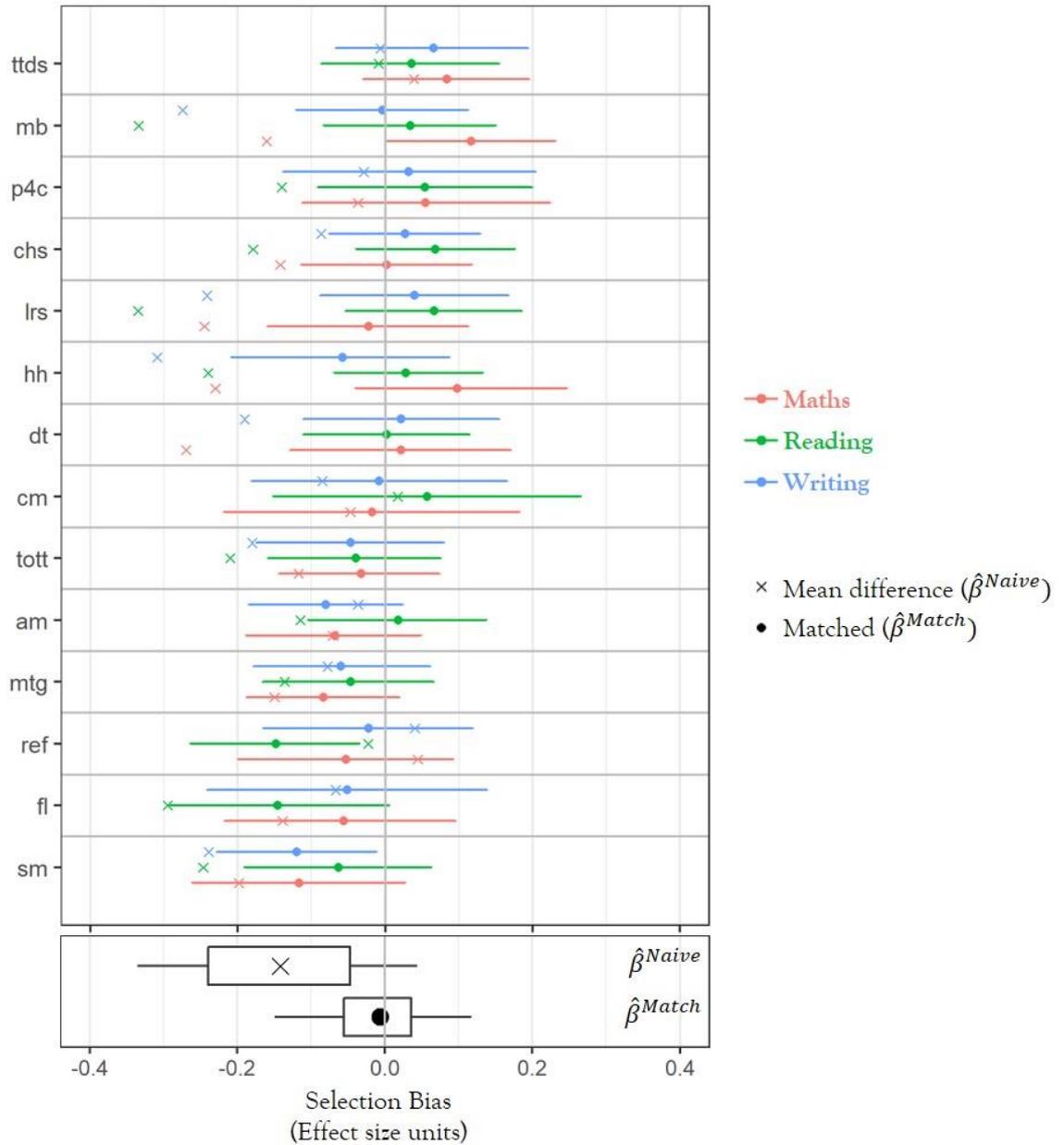

Figure 1 – raw estimates of selection bias ($\hat{\beta}^{Match}$ and $\hat{\beta}^{Naive}$)

In this section, we use a meta-analysis framework to explicitly account for sampling variation. This addresses two overlapping goals: to present estimated distributions for $\beta^{Match}$ and $\beta^{Naive}$ that are not over-dispersed due to sampling error, and to estimate the typical value of underlying selection bias for our setting. In both cases, the aim is to support researchers, bureaucrats and educators who are interested in assessing the likely magnitude of selection bias.

Observed bias estimates $\hat{\beta}_{kw}$ are assumed to be made up of several components:

$$\hat{\beta}_{kw}|\beta_{kw} \sim N(\beta_{kw}, \sigma_{kw}^2)$$
$$\beta_{kw} \sim N(\nu, \tau^2)$$

Where:

- $\nu$ = the mean bias across all interventions and outcomes
- $\beta_{kw}$ = the underlying selection bias for outcome $k$ in intervention $w$. This has a variance of $\tau^2$ reflecting the fact that not all interventions will have the same selection bias. $\beta_{kw}$ might change due to context, the nature of the program, the outcome, and so on.
- Observed bias $\hat{\beta}_{kw}$ deviates from underlying bias $\beta_{kw}$ with a variance of $\sigma_{kw}^2$. The magnitude of sampling variation largely depends on how many schools participated in intervention $w$.

Appendix F provides details of our approach to estimating these parameters, which draws heavily on random effects meta-analysis (Higgins, Thompson, & Spiegelhalter, 2009). For each intervention-outcome pair we calculate a constrained empirical Bayes' estimate of selection bias, $\tilde{\beta}_{kw}$ (Weiss et al., 2017).[12] Estimates for $\tilde{\beta}^{Match}$ and $\tilde{\beta}^{Naive}$ are presented in Figure 2.

The bottom panel of Figure 2 represents the core output of this study: this is our best guess at the distribution of underlying selection bias due to unobserved characteristics. The mean of the estimated $\tilde{\beta}^{Match}$ distribution is $\hat{\nu}^{Match} = -0.007\sigma$, indicating that across studies and outcomes the average bias tends to be very close to zero. More importantly, the mean absolute value of

---

[12] The empirical Bayes estimates are rescaled so that the empirical distribution of $\tilde{\beta}^{Match}$ ultimately has the variance $\hat{\tau}^2$. Failure to constrain the estimates in this way would lead to a distribution that was too narrow (Weiss et al., 2017).

$\tilde{\beta}^{Match}$ is $0.03\sigma$, suggesting that the typical magnitude of bias is relatively small. Last, we note that all the estimates of $\tilde{\beta}^{Match}$ are less than $0.11\sigma$.

Figure 2 - estimated distribution of $\tilde{\beta}^{Naive}$ (top) and $\tilde{\beta}^{Match}$ (bottom)

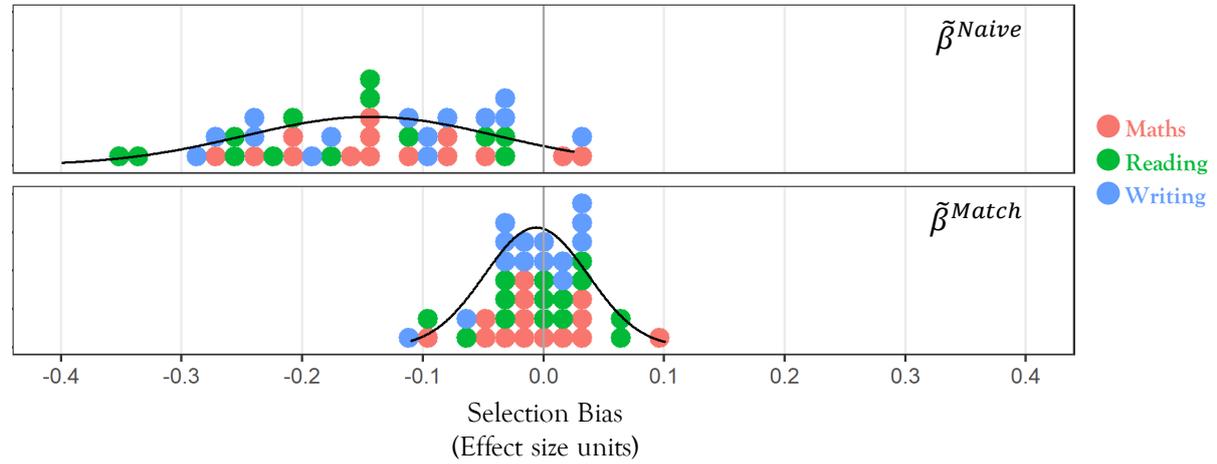

**5.3 Null Hypothesis Testing**

Figure 2 presents our final point estimates of bias. However, these estimates – and the associated distributions of underlying bias – contain uncertainty. In this section, we test the null hypothesis that selection bias is zero across all the studies and outcomes we examine, using a non-parametric inference procedure.[13] The core idea of this test is to replicate our entire analysis many times in a simulated world where, by design, the only differences between RCT control schools (CT) and comparison schools (CO) is due to sampling variation.

Broadly speaking, each simulation works as follows. For every intervention we set aside the RCT control group and replace it with a 'null control group'. The null control group is chosen at random from the pool of potential comparison schools.[14] We can then estimate $\hat{\beta}_{NULL,kw}^{Naive}$ as a difference in the mean outcome of two groups: the null control group and the pool of potential comparison schools. To generate $\hat{\beta}_{NULL,kw}^{Match}$, we find a match for each null control school and fit the multi-level model described in section 4. We complete this procedure 500 times for each intervention and outcome. This ultimately yields 500 distributions of $\hat{\beta}_{NULL}^{Match}$ and 500 for $\hat{\beta}_{NULL}^{Naive}$. We summarize each of these $\hat{\beta}_{NULL}$ empirical distributions by looking at their mean $\hat{\mu}$ and standard deviation $\hat{\sigma}$. We finally compare our actual observed estimate to these distributions, calculating how often we see a more extreme value. This is our p-value (Davison & Hinkley, 1997).

---
[13] Bias estimates within each intervention are correlated in a complex way. This motivated our use of a simulation-based approach to inference.
[14] This 'treated group' is the same size as the actual treated group for each intervention.

We find evidence to reject the null that $\hat{\beta}^{Naive}$ estimates are all zero. The observed mean ($\hat{\mu}^{Naive} = -0.15\sigma$) and the observed spread ($\hat{\sigma}^{Naive} = 0.12$) are very unlikely under the null, with p<0.01. In contrast, we find no evidence that that the observed values of $\hat{\mu}^{Match}$ and $\hat{\sigma}^{Match}$ are inconsistent with their null reference distributions. In other words, the distribution of $\hat{\beta}^{Match}$ that we observe seems to fit with the distributions we generate in a simulated world in which there is no selection bias.

Finally, as a sensitivity check, for each intervention we averaged $\hat{\beta}^{Match}$ estimates across outcomes, and tested for variation using the Q-statistic. Once again, we found no evidence for selection bias in our sample of interventions (see Appendix G for details).

### 5.4 What Predicts Individual Estimates of Bias ($\widehat{\boldsymbol{\beta}}^{Match}$)?

Although we find limited evidence of selection bias when looking at the distribution of $\beta^{Match}$ across interventions and outcomes, it may be the case that particular interventions or outcomes *are* shaped by factors we don't observe. In other words, our estimated bias distribution may be a mixture of distributions. For example, the CIA is arguably most plausible with math as an outcome. Compared to language-based outcomes, math is often thought to be less influenced by out-of-school factors and more influenced by what happens in schools (see evidence from Angrist, Dynarski, Kane, Pathak, & Walters, 2010; Dobbie & Fryer Jr, 2013; Fryer, 2014). Given that our administrative data contains more information about schools than out-of-school factors, the risk of omitted variable bias may be less severe for analyses that focus on math.

The possibility that some types of observational studies could be more prone to bias than others motivates our analysis in this section. We examine whether we can predict the magnitude of estimated bias $|\hat{\beta}_{kw}^{Match}|$ based on observable features of non-experimental studies.[15] We consider three possible predictors: outcome type (math, reading and writing), sample size, and residual imbalance in observed covariates.

First, we examine outcome type. As noted above, it is plausible that studies using math as an outcome may be less prone to selection bias compared to those focused on language-based outcomes. However, Figure 1 suggests that, at least with our straightforward matching approach, the distribution of selection bias for math is similar to that of reading and writing. The lack of association between outcome type and $\hat{\beta}$ can be tested with a simple linear regression:

$$|\hat{\beta}_{kw}^{Match}| = \alpha_0 + \alpha_1 READ_{kw} + \alpha_2 WRITE_{kw} + \epsilon_{kw}$$

where $READ_{kw}$ and $WRITE_{kw}$ are binary variables indicating that our analysis focused on intervention $w$ and outcome $k$. The results, presented in Appendix H, provide no evidence that outcome type predicts the magnitude of bias. Second, we examine whether sample size predicts estimates of bias. Again, we find no evidence of an association. Last, we examine whether the magnitude of estimated bias is associated with residual imbalance in observed covariates. For each matched dataset we summarize residual imbalance in observables by counting the number of balance violations, defined as instances where a covariate had an adjusted mean difference between treatment and control of greater than $0.25\sigma$ (Hallberg, Cook, Steiner, & Clark, 2016; Rosenbaum & Rubin, 1985). We find no evidence of an association. This suggests that the common interpretation of $\hat{\beta}^{Match}$ – that it is driven by unobserved characteristics, rather than remaining imbalances in observable characteristics – holds in our setting.

---

[15] We also applied this analysis to the constrained empirical Bayes estimates $\tilde{\beta}^{Match}$ and found the same substantive results.

## 6. Conclusion

Across 14 within-study comparisons in UK schools we did not find evidence of substantial selection bias. Our results, which significantly expand the number of cases in the literature, are broadly in line with existing estimates from evaluation of school programs in the US. The distribution of bias that we observe is centered near zero and its variance is in line with what we would expect in the absence of any selection bias. A meta-analysis of the estimates suggests that, net of sampling variation, the mean absolute value of underlying bias is 0.03σ, with a mean of -0.007σ. None of the estimates of are larger than 0.11σ. It is tempting to conclude that observational and experimental evaluations in schools will tend to produce substantively similar results. However, this conclusion goes beyond our data and is subject to several caveats.

While our analyses found limited evidence of selection bias, it doesn't follow that important unobservable factors are *generally* absent from non-experimental evaluations in schools. We searched for evidence of hidden bias in a particular set of interventions. While these programs are diverse in many ways, they shared at least three important features: they were relatively short-lived (almost all lasted less than a year); they affected only a small percentage of total instruction time; and they were all implemented in English primary schools. Unobserved characteristics may yet influence selection into more radical interventions. Broadening the set of interventions for which we have within-study comparison evidence is an important area of future work.

Similarly, it is worth reiterating that we only provide indirect evidence about selection bias in typical observational studies. We argue that the selection mechanism examined in this paper ($S_{RCT}$, which indicates whether schools select into a program and agree to be part of a rigorous evaluation) shares considerable overlap with the way that schools often select into interventions that are evaluated with observational designs ($T_{Non\_RCT}$). Although it seems unlikely to us that unobserved factors shape these two mechanisms in meaningfully different ways, this remains a possibility.

Last, we note that our results may not generalise to other times or contexts. In particular, we emphasise that the plausibility of the CIA is *always contingent* on the interaction between the nature of selection mechanisms and the data researchers can access to model those mechanisms. Both these factors are subject to change. This highlights the indispensability of RCTs, which generate unbiased estimates of sample average treatment effects and can also be used to check the ongoing trustworthiness of observational designs.

With these caveats in mind, what do our results mean for the use of experimental and observational evaluations in schools? In short, we argue that non-experimental studies can play an expanded role in building a reliable evidence base for education. First, appropriately powered observational studies should be used to identify promising programs that can then be tested with an RCT. A substantial number of programs that apply for RCT funding in a country such as England already exist in English schools. In these cases, funders could generate a local estimate of program impact by performing an observational evaluation. The results from our analyses suggest that these local observational estimates could provide valuable information about the chances that a program will succeed in a particular context. This may be particularly important given the current low success rate of interventions selected for RCTs.[16]

Second, and more broadly, we argue that researchers, bureaucrats and funders should think twice before dismissing or excessively discounting evaluations that are not based on randomised evidence. Indeed, there may well be cases in which high quality observational studies from local areas are preferable to RCT evidence from foreign or distant contexts. For example, an observational evaluation of School Peer Review currently being undertaken in the UK (see EEF, 2018) may well provide better evidence about the likelihood that the program will be effective in England than an experimental evaluation of a similar approach from a more centralized school system in another country. Observational studies – especially those that make use of large administrative datasets – can also have advantages in terms of statistical power. We have not emphasized these advantages here, as we focused on the simple bias-minimizing approach of 1:1 matching. However, other approaches to observational studies, for example full matching or inverse probability weighting, provide the opportunity for improved precision at the cost of some additional bias.

---

[16] For example, a review of education RCTs conducted from 2002 to 2012 found that 91 percent of well-implemented experiments had weak or no positive effects (CEBP, 2013).

Of course, our findings cannot disprove the basic truth that in any observational study there may be unobserved factors – "inferential monsters" in Leamer's memorable phrasing – that invalidate causal conclusions. However, in 14 cases when we were able to search for the influence of these monsters, we found little trace of them. So, while RCTs will continue to be invaluable in building a reliable evidence base for school education, well-designed observational evaluations can, and should, make more of a contribution.

**Appendix A – Evidence on Existing Within-Study Comparisons**

A review of the literature found seven interventions in school education (K-12) in which a within-study comparison had compared an RCT result to a an analysis relying on the Conditional Independence Assumption (either using matching, modelling, or both). Some of the interventions have been analysed with multiple within-study comparisons. Figure 3 presents an overview of this evidence.

Figure 3 – Summary of existing within-study comparisons in school education

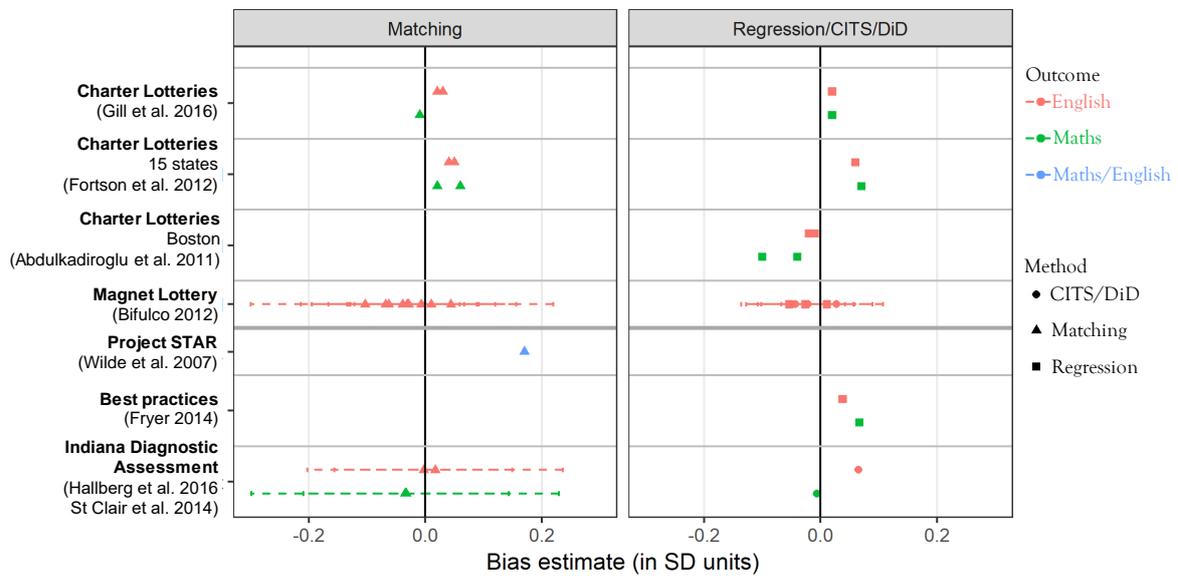

*Note*: Multiple points at the same y-axis value indicate estimates that focus on the same intervention and the same outcome, but use different analytical approaches. This Dotted lines indicate 95 percent CI for $\hat{\beta}$. Several studies do not report uncertainty estimates. CITS stands for Comparative Interrupted Time Series; DiD is the Difference-in-Difference method.

For a broader review of within-study comparisons see the excellent summary in Wong et al. (2018).

**Appendix B – Data Sources and Covariate Generation**

This section describes:

1. the different datasets we use in our analyses
2. data alterations and transformations (including trimming and imputation)
3. our approach to generating historical school-level achievement covariates

**B1 – Data**

The analysis relies on four main datasets from the UK.

**National Pupil Database**. This is a restricted database, overseen by the UK Department for Education. General information can be found in DfE (2017). The NPD amalgamates data from a variety of sources. The first source we use are the Key Stage Achievement tables. These contain information on pupil achievement in Key Stage 1 (end of Grade 2) and Key Stage 2 (end of Grade 6). In addition to pupil-level data from standardized tests, the NPD has information from a pupil census. The census is taken in autumn, spring and summer each year. It contains information on gender, student age, free-school-meal status and so on. The pupil census is complemented by a school census. This is an annual survey, sent to all schools, which yields information on the nature of school governance (e.g. whether the school is a "government" school or an "academy"), school type ("primary", "secondary") and school size. We match these censuses to two other sources: one containing school workforce information and a second on school finance. These data are collected annually by the Department, and describe staffing (e.g. pupil-teacher ratios, number of teacher assistants) and budgets. The budgetary information contains detailed financial breakdowns. In order to avoid a profusion of covariates, we pre-specified two finance variables that seemed most likely to be related to selection ($S$) and academic achievement: total per-pupil spending ("income") and per-pupil spending on educational programs (the sum of "bought in professional services", "learning resources" and ICT spending). Other variables, such as the proportion of budget spent on catering, or back-office activities, were not included in our analysis.

Moving beyond the NPD, we make use of data from **Ofsted** (the Office for Standards in Education). Ofsted provide summary evaluations of school quality, based on periodic inspections. Schools are evaluated every few years, with high-performing schools receiving less frequent inspections. In our data, schools received one of four ratings: 'outstanding=1'; 'good=2'; 'requires improvement=3'; and 'inadequate=4'. For each school, we find the most recent Ofsted evaluation before RCT recruitment. The data we use has all been published in the public domain.[17]

Third, we incorporate information on pupils' local neighbourhood by matching geographic markers in the NPD to the **English Indices of Deprivation** (DfCLG, 2015). The indices cover a wide range of characteristics at the Lower Super Output Area level – a census unit averaging about 1,500 households each. These characteristics include measures of employment deprivation, health and disability, crime, education attainment, and so on. Examination of these data revealed a very high correlation between several of these indices. To deal the possible issues of collinearity, we focussed on a subset of variables for our analysis.[18] The first is the Income Deprivation Affecting Children Index (IDACI), which is based on the proportion of children aged 0-15 living in households that are income deprived – for example households receiving unemployment benefits. Next, we use two indices that are not highly related to IDACI: the Crime Index, which measures the risk of crimes against people or property; the Housing Index, which measures proximity to community services, and housing affordability.

Last, we use the **Education Endowment Foundation RCT Archive.** Data from every completed EEF-funded evaluation is added to the archive. Most of these trials contain unique pupil IDs which can be matched to the NPD.

---

[17] We are grateful to Dave Thomson at the Education Datalab who provided us with a historical table of Ofsted evaluations for all schools.
[18] We focus on keeping a small number of neighbourhood characteristics that are not highly correlated and are conceptually distinct (Steiner, Cook, Li, & Clark, 2015)

## B2 – Alterations To the Data

Starting with full NPD cohorts, we discarded records with any of the following: a duplicated ID in EEF archive; a missing treatment indicator in the EEF archive; missing outcome data; a geographic category (e.g. "very rural") or school type (e.g. "special school") not present in our RCT sample. Next, we imputed missing data via the *mice* package (Buuren & Groothuis-Oudshoorn, 2010). Here, we took the simple approach of using the first imputed dataset from the *mice* defaults. The rate of missingness is very low in these administrative datasets. Last, inspection of the raw distributions of the budget covariates (income, outside budget) revealed extremely right skewed distributions. This prompted us to log transform these variables.

## B3 – Historical Performance Data At the School Level

This section describes how we defined and calculated measures of school-level achievement in the pre-trial period. We considered two kinds outcomes at the end of primary school:

1. scaled scores (for math and reading);

2. whether students were achieving at the expected level or above (for math and reading)

Let the year that randomisation took place for a particular RCT be $t^*$. We fit the following multilevel model for each outcome, where $Y_{ijt}$ represents an outcome for student $i$ in school $j$ at time $t$:[19]

$$Y_{ijt} = \beta_{0j} + \beta_{1j}t + \beta_{2j}t^2 + \epsilon_{ijt} \quad for\ t < t^*$$
$$\beta_{0j} = \gamma_{00} + u_{oj}$$
$$\beta_{1j} = \gamma_{10} + u_{1j}$$
$$\beta_{2j} = \gamma_{20} + u_{2j}$$

Our pupil-level data extended back to 2008-09. Schools needed at least three years of data to fit a growth curve, so we remove any observational control school that did not have at least this much data.[20] Using our model, we generated predicted values for each school of the mean value in $t^* - 1$ and the annual growth rate from $t^* - 4$ to $t^* - 1$.[21]

---

[19] To help with the estimation, all variables are scaled with mean=0 and standard deviation=1.

For each school, we ultimately generated three covariates (as per Table 1):[22]

- School academic level: estimated average of predicted math points, and predicted read points, in $t^* - 1$.
- School academic growth: estimated average annual change in reading and math points, from $t^* - 4$ to $t^* - 1$.
- School grade level growth: estimated average annual change in the combined proportion of children at grade level in reading and math from $t^* - 4$ to $t^* - 1$.

Future work should make more use of these historical data. In particular, there is an opportunity to us within-study comparisons to explore the merits of time series and Synthetic Control approaches, building on the work of St. Clair et al. (2014).

---

[20] If there are any RCT schools that don't have enough data to fit these models, we just use the raw mean outcome for t*-1 (or the closest time period). If these schools have 2 years of data, we estimate the annual growth rate by finding the difference in the mean outcome across the two years, and assuming a constant change.
[21] We use a four-year window of data in part because a within-study comparison that looked at using lagged dependent variables in observational studies suggested that going back five or six years can be problematic (St. Clair et al., 2014). There was also a practical constraint, in that our achievement data begins in 2008-09, and so for the early trials four years is the maximum possible, assuming we were to use the same covariates for all interventions.
[22] We initially considered treating reading and maths separately. However, initial analyses suggested that these variables were very highly correlated, so we decided to combine them. We also considered the average percentage of children achieving at grade level in $t^* - 1$. However, this was also highly correlated with other covariates, so we decided not to include the variable in our analysis. All these decision were taken before the bias analysis began.

**Appendix C – Propensity Score Estimation and Balance**

Our approach to estimating propensity scores is as follows. Let $S_{jw}$ be a binary indicator of whether school j selects into intervention $w$. We take simple, automated steps to specify a propensity score model that will result in satisfactory covariate balance. In doing this, our aim was to mimic the process that analysts typically use in observational studies.

1. We start with a baseline model with main effects[23]

$$\Pr(S_{jw} = 1 | X_j^M) = logit^{-1}(X_j^M \boldsymbol{\beta})$$

   where $X^M$ is a matrix of the *school-level* covariates that we use for matching. All the variables in Table 1 are included in $X^M$. Student-level covariates, such as gender, are turned into school-level variables by taking the school mean.[24] We scale all non-binary variables across the population to have mean=0 and var=1.

2. We then specify a model with interactions. To save computational time, we don't estimate a fully-interacted model. Instead, we use the baseline model to identify variables that play a role in $S_{jw}$ [i.e. those with a significant coefficient at $\alpha = 0.05$] and interact these with all other variables.

3. Last, to see if there are important non-linearities, we specify a generalised additive model (with the interactions specified in Step 2). Here, we include splines for the continuous variables listed in Table 1, including school averages of binary data for students.

For each of these three models, we generate matched datasets using the Matching package (Sekhon, 2011). We choose the model that minimises the number of variables that remain imbalanced in our matched dataset, in the sense of having a mean adjusted difference between treatment and control of more than $0.25\sigma$ (Hallberg, Wong, et al., 2016; Rosenbaum & Rubin,

---

[23] When we estimate propensity scores, we include all RCT schools in the estimation to increase precision. After estimating propensity scores, we remove the experimental treatment schools from the analysis.
[24] School means are defined by the cohort we're examining. For example, if the RCT involves grade 5 pupils, the school-average gender is the proportion of grade 5 children who are female.

1985). Given the large pool of potential controls, it's rare to have imbalance even with a large number of covariates. Table 3 provides a summary.

Table 3 – Summary of mean differences > $0.25\sigma$

| Intervention | Violations | # of covariates |
|---|---|---|
| Affordable maths | 1 | 26 |
| Changing mindset inset | 6 | 22 |
| Chess | 0 | 24 |
| Dialogic teaching | 0 | 22 |
| Flipped learning | 0 | 23 |
| Hampshire hundreds | 0 | 21 |
| Learner response system | 0 | 22 |
| Magic breakfast | 0 | 26 |
| Mind the gap | 0 | 24 |
| P4c | 2 | 23 |
| Reflected | 1 | 24 |
| Shared maths | 0 | 24 |
| Talk of the town | 0 | 25 |
| Thinking talking doing science | 1 | 22 |
| TOTAL | 11 | 328 |

*Note*: the number of covariates for each trial differs. If a trial has no children from a particular school type (e.g. "academy") or rurality classification (e.g. "very rural") then we remove the indicator variables from our analysis.

**Appendix D – Example Within-Study Comparison**

This section presents details of an example case: the *Chess in Schools* intervention evaluated by Jerrim et al. (2016). The program involved teaching chess to students in grade 5, instead of music or PE, over a period of 30 weeks. There were 100 schools in the RCT: 50 randomised to treatment and 50 to control. Recruitment for the study happened at the end of the 2012-13 academic year, and randomisation took place in July 2013. Chess was taught to students in treated schools from October 2013 to July 2014. The cohort of interest is all grade 5 pupils in English government-funded schools in 2013-14 (excluding special schools). This group sat their standardised "Key Stage 2" exams at the end of grade 6 in May 2015 (10 months after the intervention finished) and attended 13,959 schools.[25]

---

[25] All schools that participated in an EEF RCT were removed from the sample. We also removed special schools from the pool of potential comparison schools.

The first step in our analysis is to generate naïve estimates of selection bias. This is the difference in mean outcomes for the 50 experimental control schools, and the mean outcome across all 13,859 potential comparison schools.

Next, we generate covariates for the cohort of interest. We use the data sources described in section 4.1. For student-level covariates we use information in the NPD taken from before the randomisation.[26] The primary source is the Spring census of 2012-13.[27] Next, we generate school-level covariates. These data are taken from 2012-13. For example, we estimate average performance on grade 6 exams in 2012-13 (school_academic_level) and the average change in performance on grade 6 exams from 2009-10 to 2012-13 (school_academic_growth).

At this point, we have covariate information on 50 RCT schools (all those in the control condition) and 13,859 potential comparison schools. We estimate three propensity score models at the school level, as per Appendix B, and pick the propensity score model that yields the matched dataset that exhibits the best balance. Specifically, we find the model that results in the fewest variables with adjusted mean differences between treatment and control of greater than $0.25\sigma$ (Hallberg, Wong, et al., 2016; Rosenbaum & Rubin, 1985).[28] Balance is summarised in Figure 2. Our matched data has 100 schools, comprising 3,794 students. 50 of these schools are experimental controls (CT) and 50 are observational controls drawn from the pool of 13,859 potential matches. Last, we generate three estimates of selection bias – one each for math, reading and writing – using the model described in section 4.4.

---

[26] If time-invariant data such as gender is missing, we also search in Censuses that were taken after randomisation.
[27] If data is missing, we then consider the Autumn census of 2012-13.
[28] In the case of ties, we choose the simpler model.

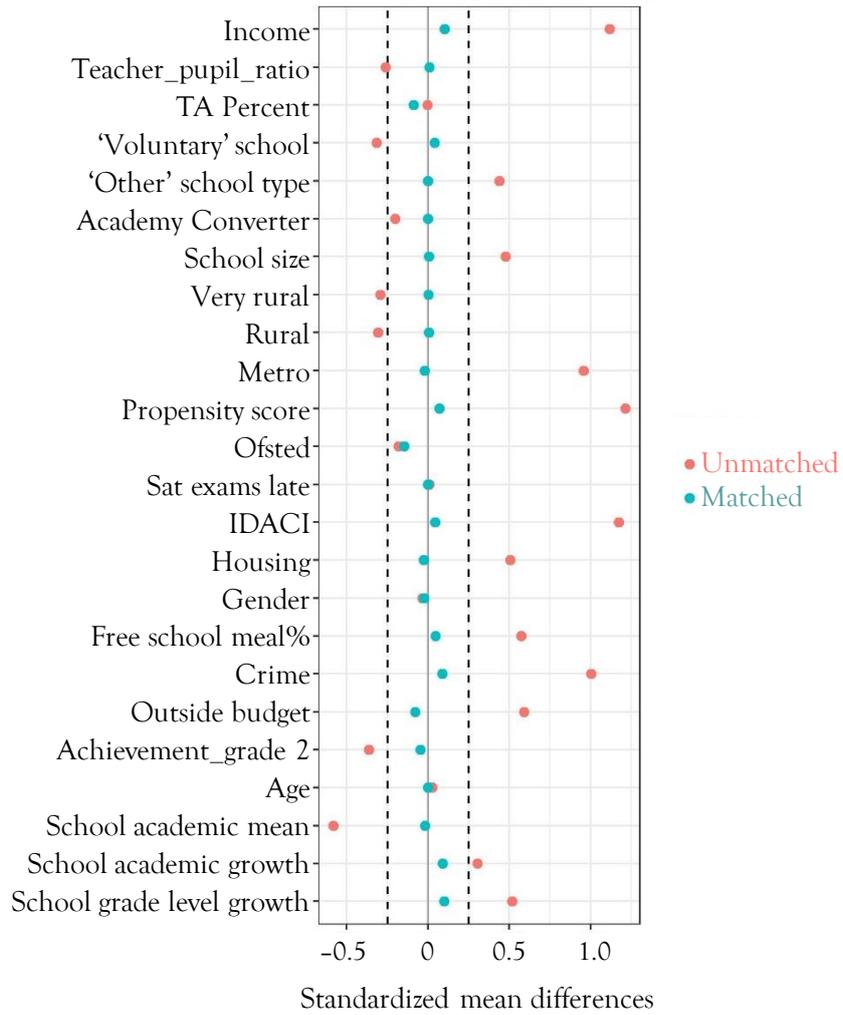

Figure 4 – Love plot for Chess in Schools analysis

*Note*: standardized mean difference is defined as the mean for the experimental control group, less the mean in the matched control group, divided by the pooled standard deviation of the population (which has been trimmed to meet a common-support criterion). See Table 1 for description of variables.

## Appendix E – Implementation of Null Hypothesis Testing

This appendix explains the inference procedure discussed in Section 5.3. First, it describes our approach to testing the null that $\hat{\beta}^{Naive}$ is zero across all interventions and outcomes. This is slightly more straightforward than our procedure for testing the equivalent null hypothesis for $\hat{\beta}^{Match}$, which is described second.

**Is There Evidence of Selection Bias In The Estimated *Naïve* Bias Distribution?**

The observed distribution of 'naïve bias' is characterised by its mean $\hat{\mu}^{Naive} = \frac{1}{42}\sum|\hat{\beta}_{kw}^{Naive}|$ and spread $\hat{\sigma}^{Naive} = \sqrt{\frac{1}{41}\sum(\hat{\beta}_{kw}^{Naive} - \hat{\mu}^{Naive})^2}$. We test the null hypothesis that none of the interventions suffer from selection bias i.e. $E[Y(0)|S=1] = E[Y(0)|S=0]$. We generate a reference distribution for two parameters under the null: $\hat{\mu}_{Null}^{Naive}$ and $\hat{\sigma}_{Null}^{Naive}$. Estimates of bias within each project are correlated. To capture these dependencies in our inference, we complete the following non-parametric process for each intervention $w$:

1. Set aside all schools with $S_{jw} = 1$
2. Repeat the following 500 times:
   a. Let $n_w$ be the number of experimental control schools in evaluation $w$. Draw a random sample of size $n_w$ from the pool of potential comparison schools. Call this sample CT†.
   b. Generate three estimates of $\hat{\beta}_{Null}^{Naive}$ (one each for math, reading and writing) by comparing the mean outcome in CT† to the overall mean outcome in the comparison pool.
   c. Return the sample CT† to the pool of potential comparison schools.

At the end of this process, we have 500 sets of $\hat{\beta}_{Null}^{Naive}$. Each set has 42 estimates of bias from across our 14 interventions and can be thought of as a single draw, *under the null*, of the bias distribution for a naïve approach to evaluation. From these 500 sets of $\hat{\beta}_{Null}^{Naive}$ we generate references distributions of $\hat{\mu}_{Null}^{Naive}$ and $\hat{\sigma}_{Null}^{Naive}$. The results of this procedure are presented in Figure 5. Both panels provide evidence against the null. The observed value for mean naïve bias is $\hat{\mu}^{Naive} = -0.15\sigma$, which is more extreme than any of the draws under the null. Similarly, the spread of our observed bias distribution $\hat{\sigma}^{Naive} = 0.12\sigma$ is also not consistent with would expect in the absence of any selection bias (p=0.006). This suggests that the sum of $\Delta_X + \Delta_U$ are non-zero.

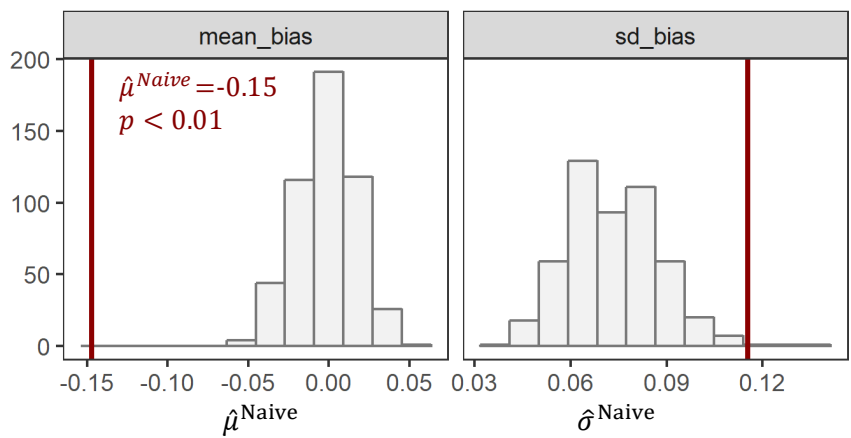

Figure 5 – Inference for 'Naïve' bias distribution

*Note*: Based on 500 simulations under the null.

**Is There Evidence of Selection Bias After Controlling for Observables?**

Having found evidence of selection bias with our naïve estimates, we see whether this changes after conditioning on observables. Now we test the null of 'no **hidden** bias'. We interpret this test as largely focussing on $\Delta_U$. We follow a similar procedure to the one described above. In this case, however, we also condition on observed characteristics, both in the matching step and then with a regression model during the estimation of $\hat{\beta}^{Match}$.

The parameters of interest are $\hat{\mu}_{Null}^{Match}$ and $\hat{\sigma}_{Null}^{Match}$, which are analogous to $\hat{\mu}_{Null}^{Naive}$ and $\hat{\sigma}_{Null}^{Naive}$. We generate reference distributions for these parameters using the following procedure. For each intervention $w$:

1. Set aside all schools with $S_{jw} = 1$
2. Repeat the following 500 times:
   a. Let $n_w$ be the number of experimental control schools in evaluation $w$. Draw a random sample of size $n_w$ from the pool of potential comparison schools. Call this sample CT*
   b. Find matches for CT* in the population of observational control schools using the matching procedure described in section 4.3. Call these matches CO*.
   c. Use the sample of CT* and CO* to estimate bias for each outcome (math, reading, writing). We use the model described in section 4.4 (replacing $S$ with $S^*$, where $S^* = 1$ if schools are in the set CT*).

This yields 500 sets of $\hat{\beta}_{Null}^{Match}$, each of which has 42 estimates of bias (in a situation where, by design, underlying bias is zero). These 500 sets of estimates are used to generate references distributions of $\hat{\mu}_{Null}^{Match}$ and $\hat{\sigma}_{Null}^{Match}$. The results of this procedure are presented in Figure 6. The left panel shows that we have no evidence to reject the null on the basis of our estimated mean bias. The mean observed estimate of $\hat{\beta}^{Match}$ is $\hat{\mu}^{Match} = -0.01\sigma$. This is consistent with the null. Similarly, the right panel suggests that the observed standard deviation of $\hat{\beta}^{Match}$ ($\hat{\sigma}^{Match} = 0.07\sigma$) is in line with what we would expect under the null. In summary, looking across 14 interventions and three domains of learning we find limited evidence that unobserved factors play a substantial role in biasing non-experimental estimates, after conditioning on covariates.

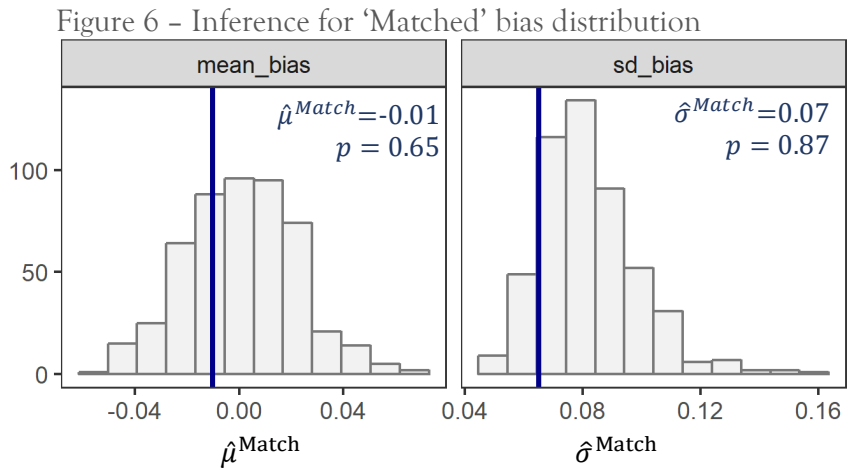

Figure 6 – Inference for 'Matched' bias distribution

## Appendix F – Meta-analysis Details

This appendix presents our approach to generating constrained empirical Bayes' estimates of bias. We use a meta-analysis framework to model sampling variation. Our observed bias estimates $\hat{\beta}_{kw}$ are assumed to be made up of several components:

$$\hat{\beta}_{kw}|\beta_{kw} \sim N(\beta_{kw}, \sigma_{kw}^2)$$
$$\beta_{kw} \sim N(\nu, \tau^2)$$

Where:
- $\nu$ = the mean bias across all interventions and outcomes
- $\beta_{kw}$ = the true selection bias for outcome $k$ in intervention $w$. This has a variance of $\tau^2$ reflecting the fact that selection bias can vary due to context, the nature of the program, the outcome, and so on.
- Observed bias $\hat{\beta}_{kw}$ deviates from underlying bias $\beta_{kw}$ with a variance of $\sigma_{kw}^2$. This sampling variation largely depends on how many schools participated in intervention $w$.

To estimate the variance of bias $\widehat{var}(\beta_{kw}) = \hat{\tau}^2$, we use the method-of-moments approach from Higgins et al. (2009):

$$\hat{\tau}^2 = \max\left\{0, \frac{Q - (K-1)}{\sum \hat{\sigma}_{kw}^{-2} - \frac{\sum \hat{\sigma}_{kw}^{-4}}{\sum \hat{\sigma}_{kw}^{-2}}}\right\}$$

Where:

$$Q = \sum(\hat{\beta}_{kw} - \bar{\beta})^2 \hat{\sigma}_{kw}^{-2}$$

$$\bar{\beta} = \frac{\sum \hat{\beta}_{kw} \cdot \hat{\sigma}_{kw}^{-2}}{\sum \hat{\sigma}_{kw}^{-2}}$$

Estimates of $\hat{\sigma}_{kw}^{-2}$ come from the simulations under the null, described in Appendix E: $\hat{\sigma}_{kw}^{-2} = var(\hat{\beta}_{kw})$. K is the effective sample size, and is based on the estimated intra-class correlation of the bias estimates, $\hat{\rho}$:[29]

$$K = \frac{kw}{1-(k-1)\cdot\hat{\rho}}$$

$$= \frac{42}{1-(3-1)\cdot 0.56}$$

$$= 19.9$$

The estimate of $\hat{\rho}$ comes from a multilevel model in which $\hat{\beta}_{kw} \sim N(\alpha_w, \sigma_e^2)$, $\alpha_w \sim N(\gamma_0, \sigma_a^2)$, and $\hat{\rho} = \frac{\hat{\sigma}_a^2}{\hat{\sigma}_a^2 + \hat{\sigma}_e^2}$. Letting $\hat{\omega}_{kw} = (\hat{\sigma}_{kw}^2 + \hat{\tau}^2)^{-1}$, we estimate the mean selection bias:

$$\hat{v} = \frac{\sum \hat{\beta}_{kw} \hat{\omega}_{kw}}{\sum \hat{\omega}_{kw}}$$

Next, we generate simple, parametric empirical Bayes estimates of the selection bias for intervention $w$ and outcome $k$:

$$\beta_{kw}^* = \hat{\lambda}_{kw}\hat{v} + (1-\hat{\lambda}_{kw})\hat{\beta}_{kw}$$

where $\hat{\lambda}_{kw} = \frac{\hat{\sigma}_{kw}^2}{\hat{\sigma}_{kw}^2 + \hat{\tau}^2}$.

While individual estimates of $\beta_{kw}^*$ minimize RMSE, an empirical distribution based on these estimates will underestimate the variability in bias estimates across studies and outcomes (Weiss et al., 2017). As such, we follow the procedure of Weiss et al. (2017, p.13) and scale our shrunken estimates so that their variance is equal to the estimated value of $\hat{\tau}^2$.

---

[29] Killip, Mahfoud, and Pearce (2004).

**Appendix G – Sensitivity Check: Meta-Analysing Bias at the Intervention Level**

There don't appear to be systematic differences across outcomes (math, reading, writing). Consequently, we present a simplified meta-analysis in which we focus on bias at the intervention level, removing the need to deal with bias estimates that are correlated within interventions. For each intervention, we simply average across math, reading and writing outcomes:

$$\hat{\beta}_w = \frac{1}{3}\Sigma\hat{\beta}_{kw}$$

$$\hat{\sigma}_w = \frac{1}{3}\Sigma\hat{\sigma}_{kw}$$

We follow the same method of moments approach described in Appendix F:

$$\hat{\tau}^2 = \max\left\{0, \frac{Q-(K-1)}{\Sigma\hat{\sigma}_w^{-2} - \frac{\Sigma\hat{\sigma}_w^{-4}}{\Sigma\hat{\sigma}_w^{-2}}}\right\}$$

Where:

$$Q = \Sigma(\hat{\beta}_w - \bar{\beta})^2 \hat{\sigma}_w^{-2}$$

$$\bar{\beta} = \frac{\Sigma\hat{\beta}_w \cdot \hat{\sigma}_w^{-2}}{\Sigma\hat{\sigma}_w^{-2}}$$

As we have 14 interventions, K=14. Using this approach, the estimated value of $Q$ is smaller than $(K-1)$, so the method-of-moments estimate of $\hat{\tau}^2$ at the intervention level defaults to zero. The 95 percent confidence interval of $\hat{\tau}^2$ using is [0,0.05].[30]

---

[30] To calculate this confidence interval, we use the test-inversion approach laid out in Weiss et al. (2017).

**Appendix H – Predicting Selection Bias**

Is there evidence that different outcomes have different bias distributions? We fit the following model:

$$\left|\hat{\beta}_{kw}^{Match}\right| = \alpha_0 + \alpha_1 READ_{kw} + \alpha_2 WRITE_{kw} + \epsilon_{kw}$$

$READ_{kw}$ and $WRITE_{kw}$ are binary variables indicating the outcome that we were focussing on for our analysis of intervention $w$. The results suggest that there is no difference across the three outcomes:

|  | |Estimated Bias| |
|---|---|
|  | bias |
| READ | -0.002 |
|  | (0.014) |
| WRITE | -0.014 |
|  | (0.014) |
| Constant | 0.059*** |
|  | (0.010) |
| Observations | 42 |
| R$^2$ | 0.029 |
| Adjusted R$^2$ | -0.021 |
| Residual Std. Error | 0.037 (df = 39) |
| F Statistic | 0.573 (df = 2; 39) |
| Note: | *p<0.1; **p<0.05; ***p<0.01 |

Next, we explore the association between sample size and the magnitude of bias, by fitting the following model:

$$\left|\hat{\beta}_{kw}^{Match}\right| = \alpha_0 + \alpha_1 SampleSize_{kw} + \epsilon_{kw}$$

where $SampleSize_{kw}$ is the number of pupils used to estimate bias for intervention $w$. We find no evidence of a linear association:

|                              | \|Estimated Bias\|       |
|------------------------------|--------------------------|
|                              | bias                     |
| Sample Size (000s of pupils) | -0.006                   |
|                              | (0.005)                  |
| Constant                     | 0.069***                 |
|                              | (0.014)                  |
| Observations                 | 42                       |
| $R^2$                        | 0.031                    |
| Adjusted $R^2$               | 0.006                    |
| Residual Std. Error          | 0.037 (df = 40)          |
| F Statistic                  | 1.261 (df = 1; 40)       |
| Note:                        | *p<0.1; **p<0.05; ***p<0.01 |

Finally, we explore whether there evidence that the quality of our matches is associated with bias. We fit the following model:

$$\left|\hat{\beta}_{kw}^{Match}\right| = \alpha_0 + \alpha_1 Count_{kw} + \epsilon_{kw}$$

where $Count_w$ is the count of balance violations, defined as instances where a covariate had an adjusted mean difference between treatment and control of greater than $0.25\sigma$ (Hallberg, Wong, et al., 2016; Rosenbaum & Rubin, 1985). As illustrated in Appendix B, this was rare. In any case, we found no evidence of a linear association between residual $X$ imbalance (defined by $V_w$) and the magnitude of bias:

|                              | \|Estimated Bias\|       |
|------------------------------|--------------------------|
|                              | bias                     |
| Residual Imbalance in X (Count) | -0.004                |
|                              | (0.004)                  |
| Constant                     | 0.057***                 |
|                              | (0.006)                  |
| Observations                 | 42                       |
| $R^2$                        | 0.033                    |
| Adjusted $R^2$               | 0.009                    |
| Residual Std. Error          | 0.037 (df = 40)          |
| F Statistic                  | 1.364 (df = 1; 40)       |
| Note:                        | *p<0.1; **p<0.05; ***p<0.01 |